\documentclass[onecolumn,showpacs,prb,preprint]{revtex4}
\usepackage{amssymb}
\usepackage{amsmath}
\usepackage{bm}
\usepackage{graphicx}
\usepackage{color}

\begin{document}

\title{Josephson current and $\pi-$state in ferromagnet \\ with embedded
superconducting nanoparticles}

\author{A.~V.~Samokhvalov$^{(1)}$,
A.~I.~Buzdin$^{(2)}$} \affiliation{$^{(1)}$ Institute for Physics
of Microstructures, Russian Academy
of Sciences, 603950 Nizhny Novgorod, GSP-105, Russia\\
$^{(2)}$ Institut Universitaire de France and Universit\'{e}
Bordeaux I, CPMOH, UMR 5798, 33405 Talence, France} 

\begin{abstract}
On the basis of Usadel equations we investigate
superconductor/ferromagnet/superconductor (S/F/S) hybrid systems which
consist of superconducting nanostructures (spheres, rods) embedded in
ferromagnetic metal. The oscillations of the critical current of the S/F/S
Josephson junctions with the thickness of ferromagnetic spacer between
superconducting electrodes are studied. We demonstrate that the $\pi $ state
can be realized in such structures despite of a dispersion of the distances
between different parts of the electrodes. The transitions between $0$ and $%
\pi $ states at some thickness of ferromagnetic spacer can be triggered by
temperature variation.
\end{abstract}


\maketitle

\section{Introduction}

The particularity of the proximity effect in superconductor/ferromagnet
(S/F) hybrid structures is the damped oscillatory behavior of the Cooper
pair wave function inside the ferromagnet \cite{BuzBul,BuzKupr} (for the
reviews see \cite{Buzdin-RMP05,GolubovRMP}). In some sort it is a
manifestation of the Larkin-Ovchinnikov-Fulde-Ferrell (LOFF) state induced
in ferromagnet (F) near the interface with superconductor (S). In contrast
with original LOFF, which is possible only in the clean superconductors, the
damped oscillatory S/F proximity effect is very robust and exists also in
the diffusive limit. This special type of the proximity effect is at the
origin of the $\pi$ Josephson S/F/S junction \cite{BuzBul,BuzKupr} which has
at the ground state the opposite sign of the superconducting order parameter
in the banks. Firstly the $\pi$- junction\ was observed at experiment in
Ref.~\cite{Ryazanov1} and since then a lot of progress has been obtained in
the physics of $\pi$- junctions and now they are proven to be promising
elements of superconducting classical and quantum circuits \cite{Ryazanov2}.
Different manifestations of unusual proximity effect and the $\pi$ states
have been observed experimentally in various layered S/F hybrids \cite%
{Jiang-PRL95,Sidorenko-AnnPh03,Garifullin-PRB02}. The proximity induced
switching between the superconducting states with different vorticities in
multiply connected hybrid S/F structures was suggested recently in \cite%
{Samokhvalov-PRB07,Samokhvalov-PRB09}. Theoretical studies and experiments
\cite{Courtois,Demler,Ryaz3,Faure-prb06} both demonstrated that in the
diffusive limit the spin-flip and spin-orbit scattering lead to the decrease
of the decay length and the increase of the oscillating period. In addition
the spin-flip scattering may generate the temperature induced transition
from $0$ to $\pi$ state of the junction \cite{Ryaz3,Faure-prb06} .

Naturally at the first stage of the work on the S/F/S junctions the systems
with planar geometry and well controlled F-layer thickness were considered.
However, now when the $\pi$ state proven to be very robust vs different
types of the impurities scattering \cite{Demler,Faure-prb06} (magnetic and
non-magnetic), interface transparency \cite{Baladie-prb03,Weides} it may be
of interest to address a question how the $\pi$ junction could realize in
the S/F/S systems with a bad defined thickness of F-spacer, in particular
for two superconducting particles imbedded in ferromagnet or between a flat
superconducting electrode and a small superconducting nanoparticle (such
situation could be of interest for the STM-like experiments with a
superconducting tip). This question is non-trivial because the transition
from $0$ to $\pi$ state occurs at the very small characteristic length \cite%
{Buzdin-RMP05} $\xi_{f}=\sqrt{D_f/h}$, where $D_f$ is the diffusion constant
in ferromagnetic metal and $h$ is the ferromagnetic exchange field, and the
typical values of $\xi_{f}$ are in the nanoscopic range. We could expect
that when the variation of the distance between different parts of
S-electrodes is of the order of $\xi_{f}$ the $\pi$ state would disappear.
Our calculations show that it is not the case and once again the $\pi$ state
occurs to be very robust and the transition between $0$ and $\pi$ states is
always present at some distance and also can be triggered by temperature
variation.

In this paper we present the results of a theoretical study of the
peculiarities of the proximity effect and Josephson current in S/F hybrids
which consist of superconducting nanostructures placed in electrical contact
with a ferromagnetic metal. The paper is organized as follows. In Sec. II we
briefly discuss the basic equations. In Sec. III we calculate the Josephson
current in two model hybrid S/F/S systems. The first system consists of two
superconducting rod-shaped electrodes imbedded in ferromagnet. The second
one is a S/F bilayer with a superconducting spherical particle at the
surface of the ferromagnetic layer. We examine the temperature dependence of
the critical current of S/F/S junction between the flat superconducting
electrode and the S-particle taking into account the spin-flip scattering.
For both cases the S/F interface transparency between superconducting
nanoparticles and ferromagnet is assumed to be low to prevent from
superconductivity destruction due to proximity. We summarize our results in
Sec. IV

\section{Model and basic equations}

Since the models of S/F/S junctions we are going to study consist of
superconducting particles embedded in a ferromagnetic matrix or placed on a
ferromagnetic substrate, we start from a description of the damped
oscillatory behavior of the Cooper wave function induced by such particle in
a ferromagnet.

We assume the elastic electron-scattering time $\tau $ to be rather small,
so that the critical temperature $T_{c}$ and exchange field $h$ satisfy the
dirty-limit conditions $T_{c}\tau \ll 1$ and $h\tau \ll 1$. In this case a
most natural approach to calculate $T_{c}$ is based on the Usadel equations
\cite{Usadel-prl70} for the averaged anomalous Green's functions $F_{f}$ and
$F_{s}$ for the F- and S-regions, respectively (see \cite{Buzdin-RMP05} for
details). These equations are nonlinear but can be simplified when the
temperature is close to the critical temperature $T_{c}$ or at any
temperature of the F-layer when the transparency of S/F interface is low. In
the F-region the linearized Usadel equations take the form
\begin{equation}\label{eq:1}
    -\frac{D_f}{2} \nabla^2 F_f
    + \left( \vert\, \omega\, \vert + \imath\, h\, {\rm sgn}\, \omega
        + 1 / \tau_s \right) F_f  = 0\,,
\end{equation}
where $D_{f}$ is the diffusion coefficient in ferromagnetic metal and $%
\omega $ are the Matsubara frequencies, $\omega =2\pi T(n+1/2)$, and $\tau
_{s}$ is the magnetic scattering time. We consider the ferromagnet with
strong uniaxial anisotropy, in which case the magnetic scattering does not
couple the spin up and spin down electron populations. Restricting ourselves
to the case of superconducting inclusions with cylindrical or spherical
symmetry (cylindrical rod-shaped or spherical particles of radius $R_{s}$),
one can easily  find the  following solutions of Eq.~(\ref{eq:1}), which
describe the distribution of anomalous Green's function $F_{f}$ in
ferromagnet ($r\geq R_{s}$) surrounding a superconducting cylinder
\begin{equation}\label{eq:2a}
    F_f^c(r) = A K_0(q r)\,,
\end{equation}
or superconducting sphere
\begin{equation}\label{eq:2s}
    F_f^s(r) = A \exp(-q r) / q r\,,
\end{equation}
where $K_{0}(z)$ is the Macdonald function,
\begin{equation} \label{eq:3}
    q = \sqrt{2/D_f} \sqrt{\vert\, \omega\, \vert + \imath\, h\, {\rm sgn}\, \omega + 1 / \tau_s}
\end{equation}
is the characteristic wave number of the order parameter variation in the
F-metal, and the amplitude $A$ is determined by the boundary conditions at
the S/F interface ($r=R_{s}$) \cite{Kuprianov-JETP88}:
\begin{equation} \label{eq:4}
    \sigma_s\, \partial_r F_s
    = \sigma_n\, \partial_r F_f\,, \quad
    F_s = F_f - \gamma_b \xi_n\, \partial_r F_f\,.
\end{equation}
The S/F interface between a particle and ferromagnet is assumed to be
characterized by the dimensionless parameter $\gamma _{b}=R_{b}\sigma
_{n}/\xi _{n}$ related to the boundary resistance per unit area $R_{b}$.
Here $\xi _{s(n)}=\sqrt{D_{s(f)}/2\pi T_{c}}$ is the superconducting
(normal-metal) coherence length, $\sigma _{s}$ and $\sigma _{n}$ are the
normal-state conductivities of the S- and F-metals, and $\partial _{r}$
denotes a derivative taken in the radial direction. We will assume that the
rigid boundary condition $\gamma _{b}\gg \min \{\xi _{s}\sigma _{n}/\xi
_{n}\sigma _{s},1\}$ is satisfied, when the inverse proximity effect and the
suppression of superconductivity is S-metal can be neglected \cite%
{Krivoruchko-PRB02,Bergeret-PRB04}. As a result, the pair amplitude $F_{s}(r)
$ at the S/F interface is equal to the one far from the boundary:
\begin{equation} \label{eq:5}
    F_s(R_s) = \frac{\Delta}{\sqrt{\omega^2 + \Delta^2}} =
    \frac{\Delta}{\omega}\, G_n\,,
\end{equation}
where $\Delta $ is the superconducting order parameter, and
\begin{equation} \label{eq:5n}
    G_n = \frac{\omega}{\sqrt{\omega^2 + \Delta^2}}
\end{equation}
is normal Green's function. Using the solution (\ref{eq:2a}) or (\ref{eq:2s}%
) we obtain from the Eqs.~(\ref{eq:4}) and (\ref{eq:5}) the value of $F_{f}$
at the S/F boundary $r=R_{s}$, and the amplitudes $A$ for both cases.
Finally, the expressions
\begin{eqnarray}
   & &F_f^c(r) = \frac{\Delta}{\omega}\, G_n
            \frac{K_0(q r)}{K_0(q R_s) + \gamma_b \xi_n q K_1(q R_s)} \,,
    \label{eq:6}\\
   & &F_f^s(r) = \frac{\Delta}{\omega}\, G_n \frac{R_s \mathrm{e}^{-q (r - R_s)}}
                          {r \left(1 + \gamma_b \xi_n (q + 1/R_s)\right)}\,.
    \label{eq:7}
\end{eqnarray}
describe the the damped oscillatory behavior of anomalous Green's function $%
F_{f}$ in ferromagnet surrounding the superconducting cylinder or sphere,
respectively.

The general expression for the supercurrent density is given by
\begin{equation} \label{eq:8}
    \vec{J}_s = \frac{i \pi T \sigma_n}{4 e} \sum_{\omega,\,\sigma=\pm}
        \left( \tilde{F}_f \nabla F_f - F_f \nabla
        \tilde{F}_f\right)\,,
\end{equation}
where $\tilde{F}_f(r,\omega) = F_f^*(r,-\omega)$.

\section{Critical current of junctions with superconducting particles}

\begin{figure}[tbp]
\centerline{
\includegraphics[width=0.5\textwidth]{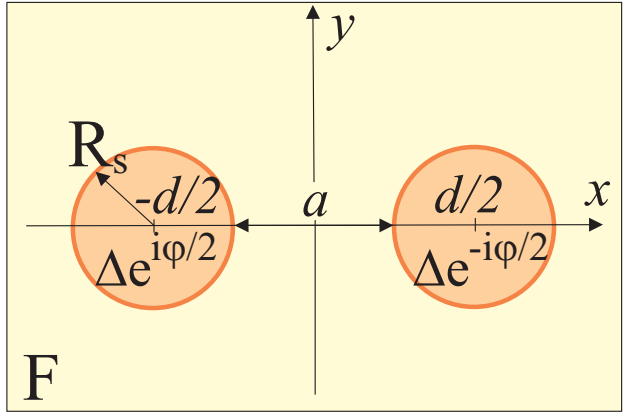}}
\caption{(Color online) Schematic representation of the F/S hybrid system
under consideration: two identical superconducting cylindrical rod-shaped
electrodes of radius $R_s$ surrounded by a ferromagnetic metal. The axes of
superconducting cylinders are assumed to be parallel. Figure shows the cross
section of the structure by the plane ($x$, $y$) perpendicular to the
cylinder axis.}\label{Fig:1}%
\end{figure}

Now we proceed with calculations of the Josephson critical current for two
examples of mesoscopic hybrid S/F systems. The first one is two identical
superconducting cylindrical rod-shaped electrodes surrounded by a
ferromagnetic metal (see Fig.~\ref{Fig:1}). The second one is a S/F bilayer
with a superconducting particle at the surface of the ferromagnetic layer
(see Fig.~\ref{Fig:3}). %
%

\subsection{S/F/S junction between two superconducting rod}

Consider two superconducting cylinders of a radius $R_{s}$ embedded in
ferromagnet as it is shown in Fig.~\ref{Fig:1}. The distance between the
cylinder axes is $d > 2R_{s}$. These rod-shaped electrodes form a Josephson
junction in which the weak link between two superconductors is ensured by
ferromagnetic neighborhood. The supercurrent
\begin{equation} \label{eq:10}
    I_s(\varphi)= I_c\,\sin(\varphi)
\end{equation}
flowing across this structure depends on the phase difference $\varphi $
between the order parameters of the rods:
\begin{equation} \label{eq:11}
    \Delta_{1,2} = \Delta\, \mathrm{e}^{\pm i \varphi/2}\,.
\end{equation}
For large enough distance between the superconducting cylinders ($%
a = d - 2R_{s} > 2\xi _{f}$),  the decay of the Cooper pair wave function in
ferromagnet in the first approximation occurs independently near either of
the electrodes and can be described by the solution (\ref{eq:6}). Therefore
the anomalous Green function $F_{f}(\mathbf{r})$ in ferromagnet nearby the
plane $x=0$ may be taken as the superposition of the two decaying functions (%
\ref{eq:6}), taking into account the phase difference $\varphi $ \cite%
{Likharev-rmp79}:
\begin{equation} \label{eq:12}
    F_f(x,y) = \frac{\Delta}{\omega}\, G_n\,
            \frac{K_0(q r_{+})\, \mathrm{e}^{i \varphi/2} + K_0(q r_{-})\, \mathrm{e}^{-i \varphi/2} }
                 {K_0(q R_s) + \gamma_b \xi_n q K_1(q R_s)}\,,
\end{equation}
where $r_{\pm }=\sqrt{(x\pm d/2)^{2}+y^{2}}$.
Using the expression (\ref{eq:8}), we obtain the sinusoidal current-phase
relation (\ref{eq:10}) in
the S/F/S Josephson junction between two superconducting rod-shaped
electrodes for the case of low transparent S/F interfaces. For the critical
current of such Josephson structure, we have
\begin{eqnarray}\label{eq:13}
    I_c &=& \frac{2\pi T \sigma_n}{e}\,
        \sum_{\omega>0} \frac{\Delta^2}{\omega^2}\, G_n^2 \nonumber \\
        &\times&\mathrm{Re}\left\{ \frac{a\, q}{\left[K_0(q R_s) + \gamma_b \xi_n q K_1(q R_s)\right]^2}
        \int\limits_{-\infty}^\infty dy \frac{K_0(q r_0) K_1(q r_0)}{r_0}
    \right\}\,,
\end{eqnarray}
where $r_{0}=\sqrt{y^{2}+d^{2}/4}$. In the limit of large $R_{s}\gg \xi _{f}$
a curvature of the electrodes is not essential and the formula (\ref{eq:13})
coincide with the corresponding expressions for the critical current of
S/F/S layered structures with a large interface transparency parameter $%
\gamma _{b}$ previously obtained in Refs.~\cite{Faure-prb06}. The critical
current equation (\ref{eq:13}) can be simplified for $h \gg \pi T_{c}$
and $R_{s},r_{0}\gg \xi _{f}$ and may be written as
\begin{eqnarray}
    I_c = I_0 \frac{d R_s}{\xi_f}
            \int\limits_{-\infty}^\infty dy\,
            \frac{\mathrm{e}^{-2(\sqrt{y^2+d^2/4}-R_s)/\xi_f}}{y^2+d^2/4}
            \cos\left(2\frac{\sqrt{y^2+d^2/4}-R_s}{\xi_f}+\frac{\pi}{4} \right),    \label{eq:14} \\
    I_0 = \frac{\pi \sigma_n \Delta\, \xi_f^2}
                 {2\sqrt{2}\, e\, \gamma_b^2\,  \xi_n^2} \tanh\left(\frac{\Delta}{2T}\right)\,.\label{eq:15}
\end{eqnarray}
%

%
\begin{figure}[tbp]
\centerline{
\includegraphics[width=0.7\textwidth]{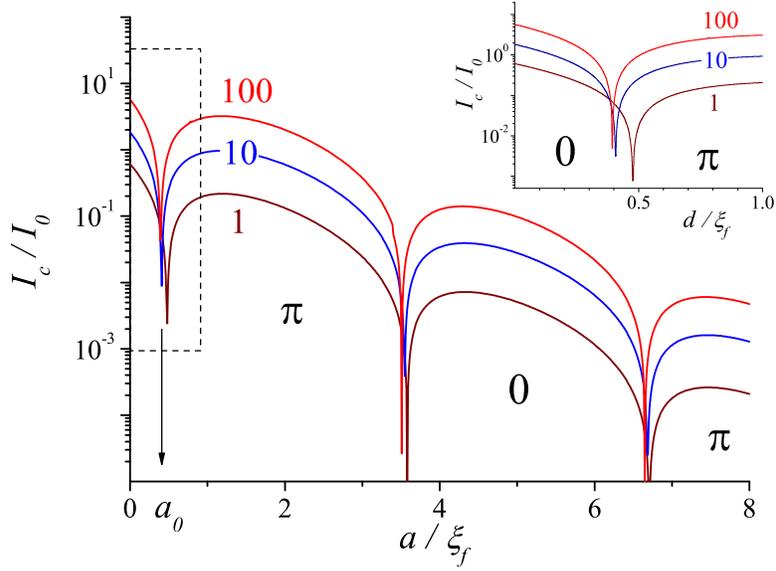}}
\caption{(Color online) Influence of the electrode radius $R_{s}$ on the
dependence of the critical current $I_{c}$ (\protect\ref{eq:14}) on the
distance $a$ between two superconducting rod-shaped electrodes embedded in
F-metal ($1/\protect\tau _{s}=0$). The numbers near the curves denote the
corresponding values of the radius $R_{s}$ in the units of $\protect\xi _{f}$%
.. The inset gives the zoomed part of the $I_{c}(a)$ line , marked by the
dashed box.}
\label{Fig:2}
\end{figure}
%

Note that our approach is valid for large enough distance between
the superconducting cylinders $a > 2\xi _f$ and the first
thansition into the $\pi$ state at $a_0$ is described only qualitatively.
It has been demonstrated \cite{Buzdin-JETPL03} that for the planar
S/F/S junction with low interface transparency the first
transition into the $\pi $ state occurs at F layer thickness
smaller than $\xi _f$ ( its actual value depends on the
the exchange field $h$ and  transparency parameter $\gamma _b$ ).%
The similar situation is expected for the embedded superconducting
particles and to find the corresponding interparticle distance we need to
solve our problem exactly.

The dependence of the critical current $I_{c}$ as a function of the distance
$a$ between the superconducting cylindrical electrodes calculated from Eq.~(%
\ref{eq:14}) is presented in Fig.~\ref{Fig:2} for several values of the
radius $R_s$. From the figure, we see that with increasing the distance $a$,
the S/F/S junction undergoes the sequence of $0$-$\pi$ and $\pi$-$0$
transitions when the value of $I_{c}$ changes its sign from positive to
negative and vice versa. We may roughly estimate that the first
transition from $0$ to $\pi $ state in S/F/S junction formed by rod-shaped
electrodes occurs at the thickness of F-layer $a_{0} \sim 0.5\xi _{f}$,
similar to a S/F/S junction with a low S/F interface transparency in the
ordinary layered geometry \cite{Baladie-prb03,Buzdin-JETPL03}.
We observe that the distances $a$ corresponding to $0$-$\pi$ and
$\pi$-$0$ transitions grow slightly with decrease of the cylinders
radius $R_{s}$ due to a dispersion of the distances between different
parts of the electrodes (see the inset in Fig.~\ref{Fig:2}).

\subsection{S/F/S junction in S/F bilayer with a superconducting
particle}

\begin{figure}[tbp]
\centerline{
\includegraphics[width=0.6\textwidth]{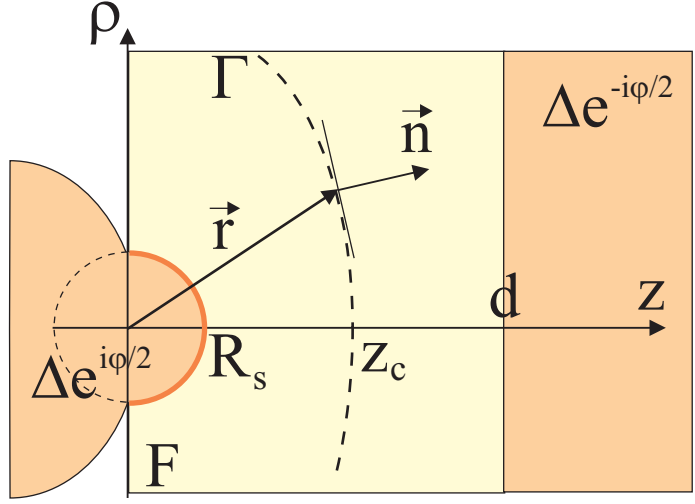}}
\caption{(Color online) Schematic representation of the S/F/S Josephson
junction between the flat superconducting electrode and a small
superconducting spherical particle of radius $R_s$ embedded into
ferromagnet. Dashed line shows the cross-section of the paraboloid $\Gamma$
by the plane ($\protect\rho,\, z$) (\protect\ref{eq:11b}): $\protect\rho%
^2=x^2+y^2$. Here $\vec{n}$ is a unit vector along the normal to the surface
$\Gamma$. Figure shows the cross section by the plane.}
\label{Fig:3}
\end{figure}

As a second example we consider ferromagnetic film of a thickness $d$ on a
superconducting plate with a transparent S/F interface. The S/F/S Josephson
junction is assumed to be formed between the flat superconducting electrode
and a small superconducting half-sphere of radius $R_s$ embedded into
ferromagnet, as it is shown in Fig.~\ref{Fig:3}. The center of the sphere is
placed at the surface of the F-film. As before, there is the tunnel barrier (%
$\gamma_b \gg 1$) at the S/F interface between the superconducting particle
and ferromagnetic metal. %
%

Since we consider a S/F bilayer with a transparent interface then the
complete nonlinear Usadel equation in the F-layer has to be employed. Using
the usual parametrization of the normal and anomalous Greens functions $%
G_{f}=\cos \Theta _{f}$ and $F_{f}=\sin \Theta _{f}$, the Usadel equation is
written as
\begin{equation} \label{eq:1b}
    - \frac{D_f}{2}\, \nabla^2 \Theta_f + \left( |\omega| + i h\, \mathrm{sgn}\, \omega
    + \frac{\cos\Theta_f}{\tau_s} \right) \sin\Theta_f = 0\,.
\end{equation}
Note that Eq.~(\ref{eq:1b}) transforms into the linear equation (\ref{eq:1})
in the limit of small $\Theta _{f}\ll 1$. For simplicity we restrict
ourselves to the case of thick F-layer ($d\gg \xi _{f}$) then the decay of
superconducting order parameter occurs independently near each S/F
interface. In that case, the behavior of the anomalous Green's function near
each interface can be treated separately, assuming that the F-layer
thickness is infinite. Following Ref.\cite{Ryaz3}~\cite{Faure-prb06}, the
analytical solution of the equation (\ref{eq:1b}) for flat transparent
interface at $z=d$ can be written as
\begin{equation} \label{eq:2b}
     \frac{\sqrt{1-\varepsilon^2\sin^2(\Theta_f/2)}-\cos(\Theta_f/2)}
          {\sqrt{1-\varepsilon^2\sin^2(\Theta_f/2)}+\cos(\Theta_f/2)}
          = f_0\, \mathrm{e}^{2 q (z - d)}\,,
\end{equation}
where
$$
    \varepsilon^2 = (1 / \tau_s) \left( \vert\, \omega\, \vert + \imath\, h\, {\rm sgn}\, \omega
        + 1 / \tau_s \right)^{-1}\,.
$$
%
The integration constant $f_{0}$ should be determined from the boundary
condition at the surface $z=d$.
As before the rigid boundary conditions is assumed to be valid at $z=d$:
\begin{equation} \label{eq:5b}
    \Theta_f(d) = \mathrm{arctan}\frac{\Delta}{\omega}\,.
\end{equation}
From Eqs.~(\ref{eq:2b}),(\ref{eq:5b}) we get
\begin{eqnarray}
    f_0 = \frac{(1-\varepsilon^2) F_n^2}
               {\left[ \sqrt{(1-\varepsilon^2) F_n^2 + 1} + 1 \right]^2 }\,, \label{eq:7b} \\
    F_n = \frac{|\Delta|}{ \omega + \sqrt{\omega^2 + |\Delta|^2} }\,. \label{eq:8b}
\end{eqnarray}
Linearizing the solution (\ref{eq:2b}) for $\Theta _{f}\ll 1$ we obtain the
anomalous Green's function in ferromagnet ($0\leq z\leq d$) induced by flat
superconductig electrode:
\begin{equation} \label{eq:9b}
    \Theta_f \simeq \frac{4 F_n}{\sqrt{(1-\varepsilon^2) F_n^2 + 1} + 1}\, \mathrm{e}^{q (z - d)}\,.
\end{equation}
The total anomalous Green's function in F-layer far from both the S/F
interfaces may be taken as superposition of the two decaying functions (\ref%
{eq:7}), (\ref{eq:9b}), taking into account the phase difference in each
superconducting electrode
\begin{equation} \label{eq:10b}
    F_f = \frac{\Delta}{\omega}\, G_n \frac{R_s \mathrm{e}^{-q (r - R_s) + i \varphi/2}}
                          {r \left(1 + \gamma_b \xi_n (q + 1/R_s)\right)}
         +\frac{4 F_n\,\mathrm{e}^{q (z - d) - i \varphi/2}}
                   {\sqrt{(1-\varepsilon^2) F_n^2 + 1} + 1}\,,
\end{equation}
where $r=\sqrt{x^{2}+y^{2}+z^{2}}$.

To derive the general expression for the critical current $I_c$ we have to
calculate the total Josephson current flowing through a virtual surface $%
\Gamma$: the points of the surface $\Gamma$ are equidistant from both
electrodes of the junction. The surface $\Gamma$ is a paraboloid which form
is described by the equations:
\begin{equation} \label{eq:11b}
    z = z_c - (x^2 + y^2) /4 z_c \,, \quad z_c = (R_s + d) / 2\,.
\end{equation}
Using the solution (\ref{eq:10b}) and Eq.(\ref{eq:8}), one can arrive at a
sinusoidal current-phase relation (\ref{eq:10}) with the critical current
(see \ref{Apx-A} for details): 
\begin{eqnarray} \label{eq:12b}
    I_c &=& I_0\, \frac{z_c R_s T}{\xi_f T_c}\, \mathrm{Re}\,\sum_{n=0}^\infty
            \left\{ \frac{ F_n\, (\Delta\,G_n/\,\omega)\, \mathrm{e}^{-q (d - R_s)}}
                         {[\sqrt{(1-\varepsilon^2) F_n^2 + 1} + 1]\,
                          [1 + \gamma_b \xi_n (q + 1/R_s)]} \right.         \nonumber \\
              &\times& \left. \left( q \int\limits_0^{z_c} du \frac{\mathrm{e}^{-2 q u}}{u + z_c}
                  + \frac{1}{2} \int\limits_0^{z_c} du \frac{\mathrm{e}^{-2 q u}}{(u + z_c)^2}
                  \right) \right\} \,,
\end{eqnarray}
where $I_0 = 64 \pi^2 T_c \sigma_n \xi_f / e$.

\begin{figure}[tbp]
\centerline{
\includegraphics[width=0.6\textwidth]{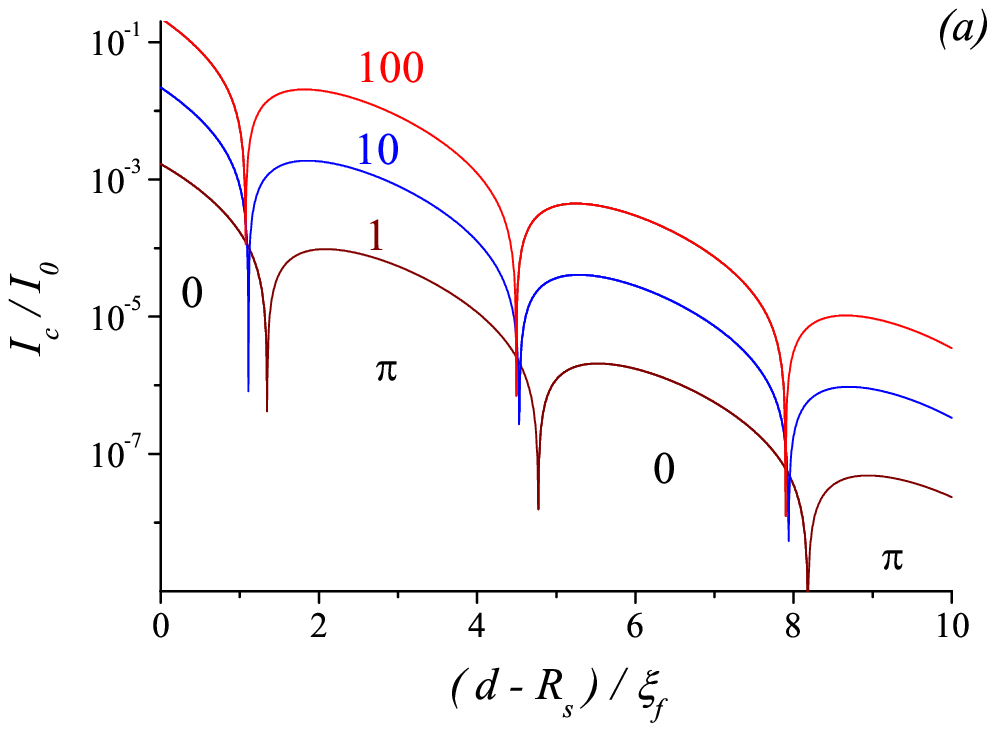}}
\centerline{
\includegraphics[width=0.6\textwidth]{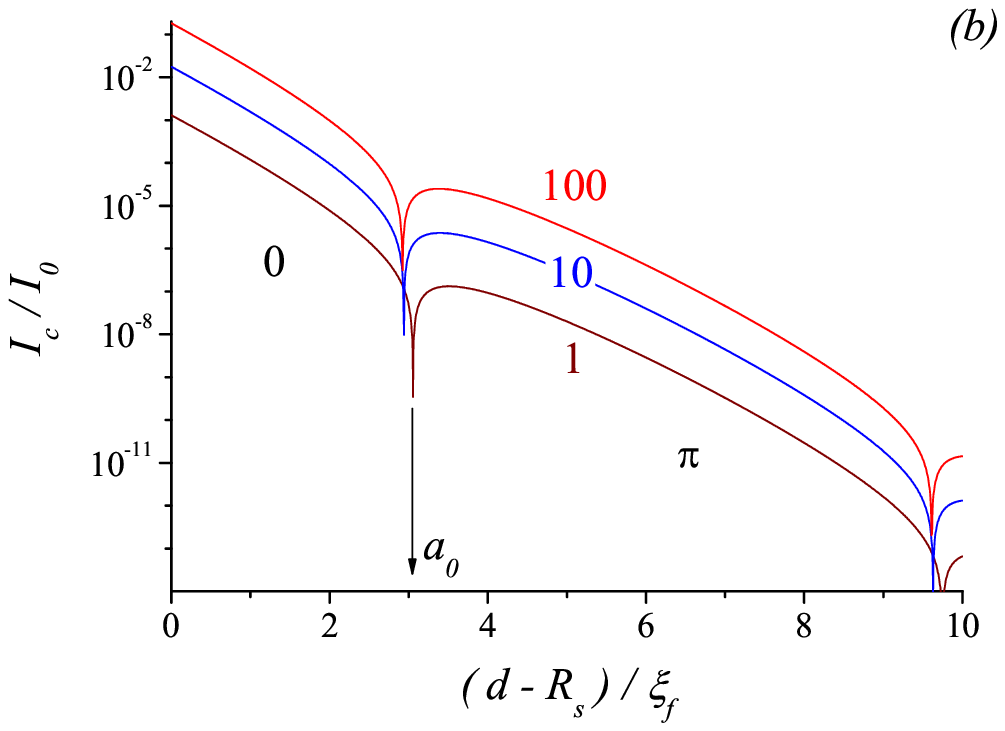}}
\caption{(Color online) The dependence of the critical current $I_c$ (%
\protect\ref{eq:12b}) on the distance $d-R_s$ between the superconducting
plate and the particle of radius $R_s$ embedded in F-metal for different
values of the radius $R_s$ and the magnetic scattering time $\protect\tau_s$
($T / T_c = 0.5$, $h = 3\protect\pi T_c$, $\protect\gamma_b = 10$): (a) $h
\protect\tau_s = 100$; (b) $h \protect\tau_s = 0.5$. The numbers near the
curves denote the corresponding values of the radius $R_s$ in the units of $%
\protect\xi_f$.}
\label{Fig:4}
\end{figure}
%

The dependence of the critical current $I_{c}$ (\ref{eq:12b}) as a function
of the thickness $a=d-R_{s}$ of the ferromagnetic spacer separating the
superconducting plate and the particle is presented in Fig.~\ref{Fig:4} for
several values of the particle radius $R_{s}$ and the magnetic scattering
time $\tau _{s}$. It is clearly seen from Fig.~\ref{Fig:4} that with a
decrease of the particle radius $R_{s}$, the position $a_{0}$ of the first
zero of the critical current is shifted towards larger values of the
distance $a$ between superconducting electrodes. Figure~\ref{Fig:4}b
demonstrates the influence of magnetic scattering on the proximity effect
and the critical current in S/F bilayer with the particle: decrease of the
magnetic scattering time $\tau _{s}$ leads to the decrease of decay length
and increase of the oscillation period of the anomalous Green's function $%
F_{f}$ \cite{Ryaz3}. This results in much stronger decrease of the critical
current in the S/F/S junction with increase of the thickness $a$, if the
magnetic scattering time $\tau _{s}$ becomes relatively small $\tau
_{s}^{-1}\geq h$.

\begin{figure}[tbp]
\centerline{
\includegraphics[width=0.6\textwidth]{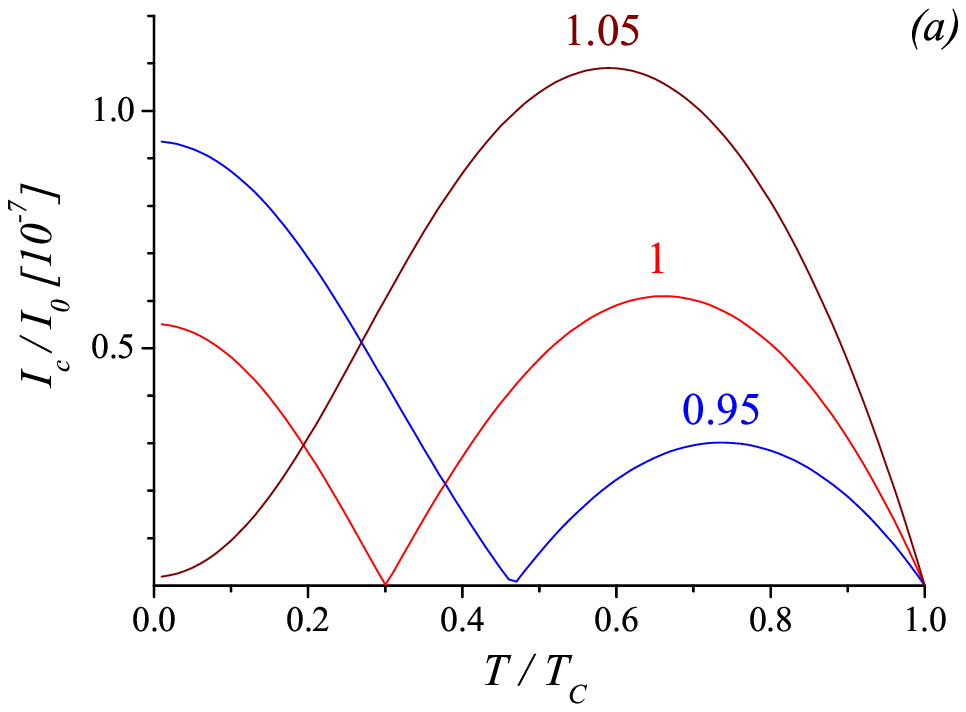}}
\centerline{
\includegraphics[width=0.6\textwidth]{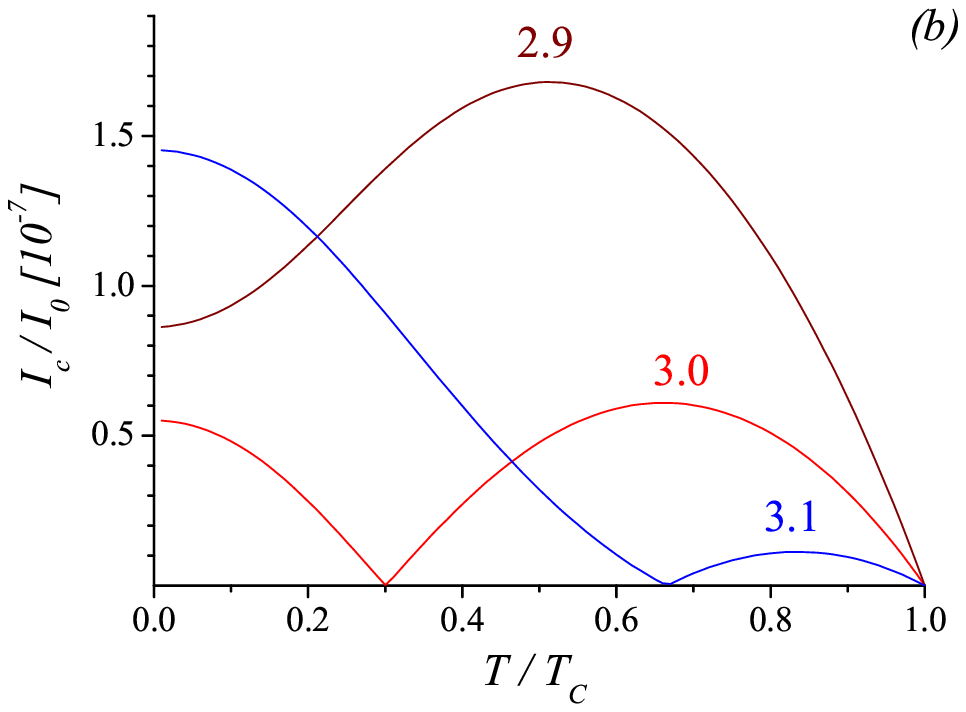}}
\caption{(Color online) (a) The dependence of the critical current $I_c$ (%
\protect\ref{eq:12b}) on the temperature $T$ for different values of the
radius $R_s / \protect\xi_f = \mathrm{0.95,\, 1.0,\, 1.05}$ and the fixed
F-spacer thickness $a = \mathrm{3} \protect\xi_f$. (b) The dependence of the
critical current $I_c$ (\protect\ref{eq:12b}) on the temperature $T$ for
different values of the F-spacer thickness $a / \protect\xi_f = \mathrm{%
2.9,\, 3.0,\, 3.1}$ and fixed radius $R_s = \protect\xi_f$. The calculation
parameters are $h = 3\protect\pi T_c$, $h \protect\tau_s = 0.5$, $\protect%
\gamma_b = 10$.}
\label{Fig:5}
\end{figure}
%
Figure~\ref{Fig:5} shows the temperature dependence of the S/F/S junction
critical current $I_c$ (\ref{eq:12b}) at several values of the thickness of
the ferromagnetic spacer between the superconducting electrodes. The
F-spacer thickness $a$ is chosen close to the first transition from $0$ to $%
\pi$ state: $a \sim a_0 \simeq 3\xi_f$. The nonmonotonic dependences $I_c(T)$
demonstrate $0$ - $\pi$ transition due to a change of the temperature. The
transition temperature $T^*$ ($I_c(T^*) = 0$) seems to be very sensitive to
the size of the superconducting particle. It should be noted however that
the temperature $T^*$ is determined rather by the scale $a = d - R_s$ then
by the scales $d$ or $R_s$, separately.

\section{Conclusion}

To sum up, we have analyzed the Josephson effect in S/F/S hybrid structures
with a bad defined thickness of F-spacer. As an example, we have calculated
the Josephson current between two rod-shaped superconducting electrodes
embedded in ferromagnet or between flat superconducting electrode and the
small superconducting nanoparticle at the surface of the F-layer. For the
both cases we have demonstrated the possibility of the realization of $\pi $
junctions in such hybrid systems. We have studied dependence of the
transitions between $0$ and $\pi $ states both on the size of
superconducting particles and the temperature. The $\pi $ state has been
proven to be very robust with respect to a geometry of the S/F/S junction.
In the dirty limit the transition into $\pi $ state is determined rather by
the thickness of the F-spacer between superconducting electrodes then by a
shape of the electrodes. Naturally our calculations can be easily
generalized to the different shape of the S particles (for example
spherical) with similar conlusions.

A set of the superconducting particles embedded in a ferromagnetic
matrix realize a Josephson network. Depending on the geometry of this
network and its state ($0$ or $\pi$) it may reveal  a
spontaneous current similar to that observed in superconducting arrays of %
$\pi$ junctions \cite{Ryaz4} .
For example the equilibrium phase difference for triangular 2D $\pi$
junctions network is equal to $2\pi /3$ which corresponds to the current
state.
For typical parameters Nb/CuNi hybrid system ($T_{c}=9\mathrm{K}$, %
$\xi _{f}\approx 2\mathrm{nm}$, %
$\rho _n=1/\sigma _n \approx 60 \mathrm{\mu \Omega \,cm}$
\cite{Ryaz3} one can get from (\ref{eq:12b})
the following estimate of the Josephson energy
$E_{J}=\phi _{0}I_{c}/2\pi c$ of the S/F/S junction:
$E_{J}/T_{c}\sim 10^{4}\,(I_{c}/I_{0})<1$, i.e. an observation
of spontaneous currents near $T_{c}$ is expected to be
masked by strong temperature fluctuations.
Despite of this restriction we believe that intrinsically--frustrated
superconducting networks induced by the proximity effect can be
experimentally observable in such S/F/S composites.
In particular, a  two-dimensional Josephson network of $\pi$%
junctions may serve as a laboratory  to study the phase
transitions with continuous degeneracy \cite{Korshunov}.

The possibility to fabricate the regular 2D and 3D arrays of
Josephson $\pi$ junctions and monitor the transitions between $0$ %
and $\pi$  states simply varying the temperature open
interesting perspectives to study a very reach physics of different phase
transition in such systems due to interplay between fluctuations,
frustration, disorder and dimensionality.

\section{Acknowledgments}
\label{acknowledgments}

We are indebted to A.~S.~Mel'nikov for useful discussions. This work was
supported, in part, by the Russian Foundation for Basic Research, by Russian
Agency of Education under the Federal Program "Scientific and educational
personnel of innovative Russia in 2009-2013", by International Exchange
Program of Universite Bordeaux I, by French ANR project "ELEC-EPR", and by
the program of LEA Physique Theorique et Matiere Condensee.

\appendix

\section{Josephson current in S/F bilayer with a superconducting
particle}\label{Apx-A}

The general expression for the supercurrent is given by the Eq.~(\ref{eq:8}),
where the anomalous Greens function $F_f$ 
nearby the surface $\Gamma$ (see Fig.~\ref{Fig:3}) may be written as
\begin{eqnarray} 
    F_f = A_1\,\mathrm{e}^{q z - i \varphi/2}
           + A_2 \mathrm{e}^{-q \sqrt{\rho^2+z^2} + i\varphi/2}\,, \label{eq:a1} \\
    A_1 = \frac{4 F_n\,\mathrm{e}^{-q d}}
               {\sqrt{(1-\varepsilon^2) F_n^2 + 1} + 1}\,,      
    \qquad A_2 = \frac{(\Delta G_n / \omega)\,\mathrm{e}^{q R_s}}
                          {1 + \gamma_b \xi_n (q + 1/R_s)}\,. \label{eq:a2}
\end{eqnarray}
where $r^2 = \rho^2 + z^2$, and the functions $G_n$, $F_n$ are determined by
expressions (\ref{eq:5n}) and (\ref{eq:8b}), respectively.
At the surface $\Gamma$ the function $F_f$ and the projection of
the vector $\nabla F_f$ along the normal
$$
    \vec{n}=\frac{\rho}{\sqrt{\rho^2+4z_c^2}}\, \vec{\rho}_0
    + \frac{2z_c}{\sqrt{\rho^2+4z_c^2}}\, \vec{z}_0
$$
to the surface are %
\begin{eqnarray}
    \left.F_f\right|_\Gamma = \left( A_1\,\mathrm{e}^{q z_c - i \varphi/2}
            + A_2 \frac{4 z_c R_s}{\rho^2+4z_c^2}\mathrm{e}^{-q z_c + i\varphi/2} \right)
                            \mathrm{e}^{-q \rho^2 / 4 z_c}\,,  \label{eq:a4} \\
    \left.(\nabla F_f, \vec{n})\right|_\Gamma = \frac{2 z_c\, \mathrm{e}^{-q \rho^2 / 4 z_c}}
                                                     {\sqrt{\rho^2+4z_c^2}} \nonumber \\
            \qquad\times\left[ q A_1\,\mathrm{e}^{q z_c - i\varphi/2}
                         - A_2 \frac{4 z_c R_s}{\rho^2+4z_c^2} \left(q + \frac{4 z_c}{\rho^2+4z_c^2}\right)\,
                        \mathrm{e}^{-q z_c + i\varphi/2} \right]\,. \label{eq:a5}
\end{eqnarray}
%
Substitution of Eqs.~(\ref{eq:a4}), (\ref{eq:a5}) into the expression for
the supercurrent (\ref{eq:8}) and taking into account the symmetry relations
$q(-\omega) = q^*(\omega)$, $A_{1,2}(-\omega) = A_{1,2}^*(\omega)$ leads to
the following formula
\begin{eqnarray}
    \left.(\vec{J}_s, \vec{n})\right|_\Gamma = J_c(\rho)\, \sin\varphi\,, \label{eq:a6} \\
    J_c(\rho) = \frac{32 \pi T \sigma_n}{e} \mathrm{Re}
        \sum_{\omega>0} \left\{\frac{z_c^2 R_s\, A_1 A_2}{(\rho^2+4z_c^2)^{3/2}}\,
        \left(q + \frac{4 z_c}{\rho^2+4z_c^2}\right)\mathrm{e}^{-q \rho^2 / 4 z_c} \right\}. \label{eq:a7}
\end{eqnarray}
Further integration of supercurrent density (\ref{eq:a7}) over the surface $%
\Gamma$
\begin{equation} \label{eq:a8}
    I_c = \int\limits_\Gamma dS\, J_c = \frac{\pi}{z_c}\int\limits_0^{2z_c} d\rho\, \rho\sqrt{\rho^2+4z_c^2}\, J_c(\rho)\,.
\end{equation}
results in the expressions (\ref{eq:12b}) for the critical current $I_c$ of
the S/F/S Josephson junction between superconducting plate and particle.

\end{document}